\documentclass[%
 superscriptaddress,
 preprint,
 showpacs,preprintnumbers,
 amsmath,amssymb,
 aps,
 prb,
]{revtex4-1}

\usepackage{graphicx}
\usepackage{dcolumn}
\usepackage{bm}

\usepackage{upgreek}
\usepackage{amsmath}

\begin{document}


\title{Strong coupling between excitons and quasi-Bound states in the continuum in the bulk transition metal dichalcogenides}

\author{ Meibao Qin }
\affiliation{Department of Physics, Nanchang University, Nanchang 330031, People’s Republic of China}

\author{Junyi Duan}
\email{jyduan@email.ncu.edu.cn}
\affiliation{Institute for Advanced Study, Nanchang University, Nanchang 330031, People’s Republic of China}
\affiliation{Jiangxi Key Laboratory for Microscale Interdisciplinary Study, Nanchang University, Nanchang 330031, People’s Republic of China}

\author{Shuyuan Xiao}
\affiliation{Institute for Advanced Study, Nanchang University, Nanchang 330031, People’s Republic of China}
\affiliation{Jiangxi Key Laboratory for Microscale Interdisciplinary Study, Nanchang University, Nanchang 330031, People’s Republic of China}
\author{Wenxing Liu }
\affiliation{Department of Physics, Nanchang University, Nanchang 330031, People’s Republic of China}
\author{Tianbao Yu }
\email{yutianbao@ncu.edu.cn}
\affiliation{Department of Physics, Nanchang University, Nanchang 330031, People’s Republic of China}
\author{ Tongbiao Wang }
\affiliation{Department of Physics, Nanchang University, Nanchang 330031, People’s Republic of China}
\author{ Qinghua Liao }
\affiliation{Department of Physics, Nanchang University, Nanchang 330031, People’s Republic of China}
\begin{abstract}
We investigate the strong coupling between the excitons and quasi-bound states in the continuum (BIC) resonance in a bulk WS$_2$ metasurface. Here we employ the bulk WS$_2$ to construct an ultrathin nanodisk metasurface, supporting the symmetry-protected magnetic dipole (MD) quasi-BIC resonance, which can self-hybridize with the excitons and lead to a strong light-matter interaction enhancement within the structure without the necessity for an external cavity. This strong coupling can be charactered by the considerable Rabi splitting of 159 meV and the clearly anti-crossing behavior appeared in the absorption spectrum. Furthermore, we analyze such light-matter coupling by constructing a Hamiltonian model including the surplus excitons, and tune the interaction from weak coupling to strong coupling regimes via the tunability radiation loss of the quasi-BIC resonance. Our results have great potential for manipulating the exciton-polaritons at room temperature, and provide a promising prospect for photonic devices that exploit strong coupling in applications.
\end{abstract}

\maketitle


\section{\label{sec1}Introduction}

Strong coupling, allowing energy to exist between light and matter in a coherent and reversible manner, has emerged as a growing range of applications such as Bose-Einstein condensation, superfluidity, spin switch, and lasing\cite{Deng2010,Kasprzak2006,Amo2009,Amo2010,Su2017}. It is demonstrated that strong coupling can be easily achieved at the room temperature by integrating the exciton in monolayer transition metal dichalcogenides (TMDCs) into plasmonic/dielectric cavities\cite{Dovzhenko2018,Zu2016,Asham2021,Anantharaman2021,Qing2022,Huang2022,AlAni2022,Xie_2022}. However, the plasmonic materials are suffered from the low damage thresholds, making them bottlenecked in further practical applications\cite{Hugall2018,Xiao2020a}. High refractive dielectric resonators support the Mie resonance without ohmic loss but confine the electric field to the internal of the structure, resulting a small Rabi splitting\cite{Kuznetsov2016,Chen2020,Koshelev2020}. Quasi-bound states in the continuum (BICs) with high Q and the ability to channeling the energy radiation to the external of the structure in dielectric metasurfaces show a prospect for overcoming this barrier. Nevertheless, the spatial overlap between excitons and quasi-BIC remain not enough to realize a large Rabi spitting\cite{Koshelev2018,Koshelev2018a,Cao2020,AlAni2021,Qin2021,Xie2021,Zhang2022,Qin2022}.
 
 Bulk TMDCs are featured by the strong exciton absorption peak and high refractive index, making it possible to the spatial overlap between the excitons and the optical resonance mode in the same structure, which provides a promising candidate for boosting strong coupling\cite{Verre2019,Wang2022}. According to the literature, very large values of Rabi splitting have been observed in some bulk TMDCs cavities, including Fabry-Perot microcavity, anapole states in a nanodisk, and quasi-BICs in a one-dimensional grating\cite{Munkhbat2018,Verre2019,Zong2021}. However, there are few researches reported in the strong exciton-BIC coupling in a bulk TMDC measurface, and in particular the surplus excitons in the coupling region and their line-width influence on the light-matter absorption spectrum need to be explored.
 
In this work, for the first time to our knowledge, we study the strong exciton-BIC interaction in a bulk TMDC measurface. We design a self-hybridized exciton-polaritons nanodisk measurface by employing the bulk WS$_2$, in which the spatial overlap between the symmetry-protected magnetic dipole (MD) quasi-BIC resonance and exciton mode can be realized, resulting in a large Rabi splitting of 159 meV. This strong coupling could be further demonstrated by the anti-crossing behavior in the absorption spectrum. Furthermore, we construct a Hamiltonian model involving the surplus excitons to interpret the coupling condition, which makes it possible to tune the absorption spectrum by adjusting the asymmetry parameter. 

\section{\label{sec2} Geometric structure and numerical model}
\begin{figure}[htbp]
	\centering
	\includegraphics
	[scale=0.5]{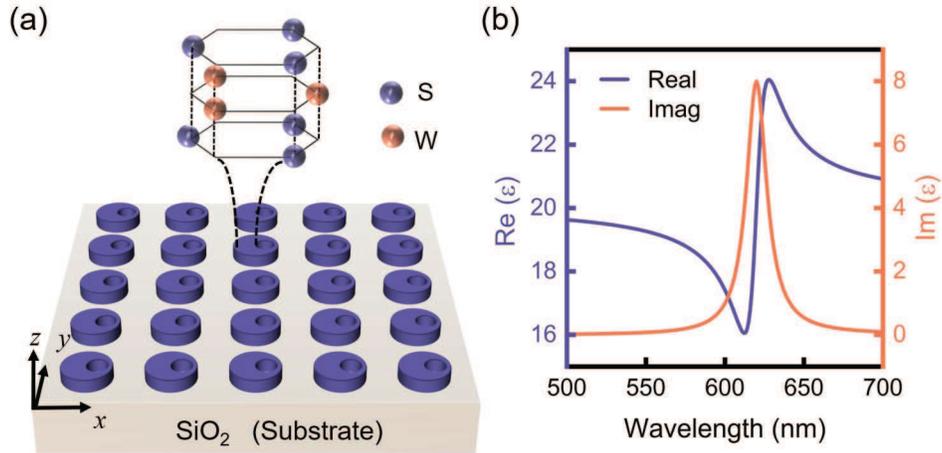}
	\caption{\label{fig1} (a)  Sketch of the system: a nanodisk metasurfaces of a bulk WS$_2$ is illuminated by a normally incident plane wave. (b) The dielectric functions of bulk WS$_2$: blue and red lines represent real and imaginary parts, respectively.}
\end{figure}

The proposed model is schematically shown in Fig. \ref{fig1}(a). A nanodisk metasurface made out of bulk WS$_2$ is deposited onto a SiO$_2$ substrate, illuminated by a normally incident plane wave. The periodicity of each unit cell is $P=370$ nm. Every disk has a radius $R=110$ nm and a height $H =35$ nm, respectively. A hole with variable radius $r$ at a fixed distance $D=R/2$ away from the center of the disk, and the asymmetry parameter $\alpha=\frac{\pi r^2}{\pi R^2}$ is characterized by the ratio of the hole area to the nanodisk area. This in-plane perturbation of the structure will open a radiation channel, enabling an energy change with the external mode by transforming the ideal-BIC into quasi-BIC.

The full-wave electromagnetic response simulation is carried out using the COMSOL Multiphysics based on the finite element method. In the simulation domain, the $y$-polarized plane wave is normally incident and the perfectly matched layers are utilized to absorb the outgoing radiated waved in the $z$ direction, where the periodic boundary conditions in the $x$ and $y$ directions due to the periodicity of nanodisk metasurfaces. For the material modeling, we choose bulk WS$_2$ as the constituent due to its high permittivity as depicted in Fig. \ref{fig1}(b). The artificial permittivity approximating of the bulk WS$_2$ in Fig. \ref{fig1}(b) can be described by\cite{Munkhbat2018}
\begin{equation}
	\varepsilon_{art} =\varepsilon_b+\frac{f_0E_{ex}^2}{E_{ex}^2-E^2-i\Gamma_{ex}E},
	\label{eq1}
\end{equation}
where $\varepsilon_b=20$ represents the background permittivity, $f_0=0.2$ is the oscillator strength, $E_{ex}=2$ eV and $\Gamma_{ex}$=50 meV are transition energy and full linewidth of the exciton, respectively. It is noticed that when we only consider it as a background index-only material $f_0=0$. For an exciton-only material, $\varepsilon_b=1$.

\section{\label{sec3} Results and discussions}

\begin{figure}[htbp]
	\centering
	\includegraphics
	[scale=0.42]{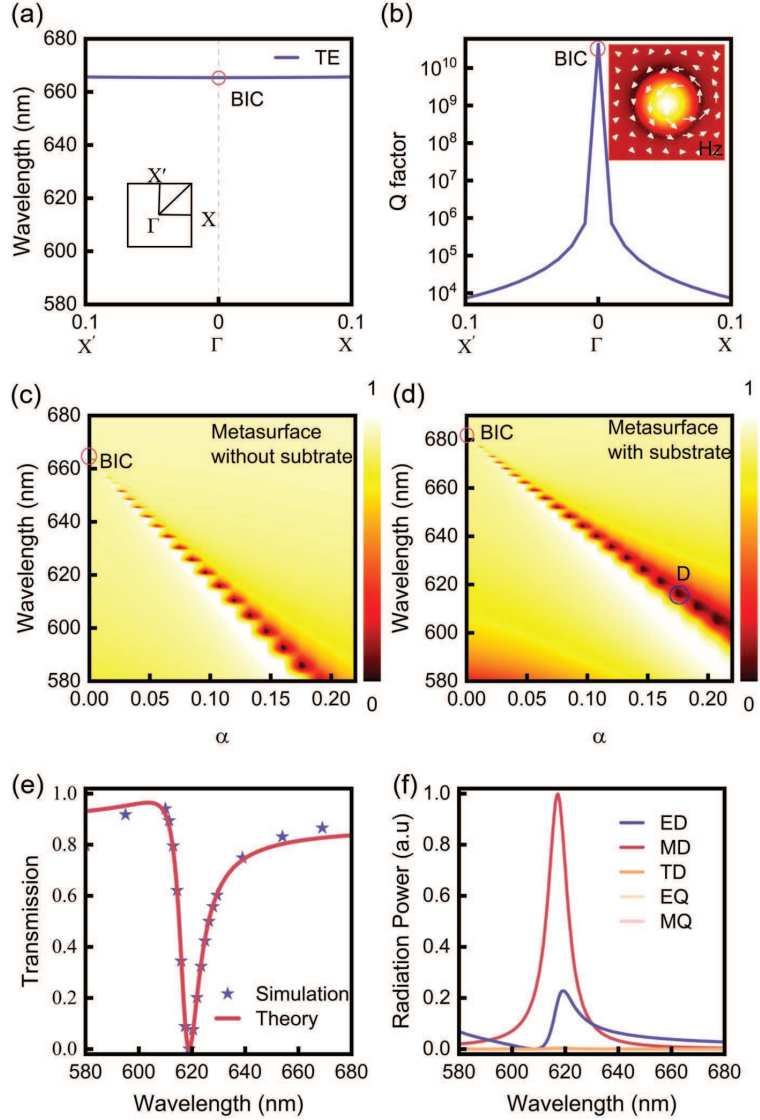}
	\caption{\label{fig2}(a) Band diagrams of the bulk WS$_2$ nanodisk metasurfaces without substrate. Here, $P=370$ nm, $R=110$ nm, and $H=35$ nm. The symmetry-protected BIC is marked with a red circle. (b) The corresponding $Q$ factors for the TE-like around the 665 nm, and the inset shows the corresponding magnetic distribution of $Hz$ at the point around the 665 nm, overlaid with the in-plane electric field direction vector $E_{xy}$.(c)The transmission spectra of bulk WS$_2$ nanodisk metasurfaces as a function of the asymmetric parameter without substrate, where an ideal BIC is marked by a red circle. (d) The transmission spectra of bulk WS$_2$ nanodisk metasurfaces as a function of the asymmetric parameter with SiO$_2$ substrate, where an ideal BIC is marked by a red circle and the D point around the exciton position of 620 nm. (e) The simulated and theoretical transmission spectra of the D point. (f) The scattered power of the multipole moments under the Cartesian coordinates.}
\end{figure}

To confirm the existence of the BIC in our proposed bulk WS$_2$ nanodisk measurface, we setting the oscillator strength $f_0=0$ and then perform an eigenmode analysis without substrate. As shown in Fig. \ref{fig2}(a), a band diagram along $X$-$\Gamma$-$X^{'}$ for transverse electric (TE)-like mode around 665 nm in the first Brillouin zone. In the following analysis of the eigenmode at $\Gamma$ point in Fig. \ref{fig2}(b), we can see that the $Q$ factor is larger than $10^{10}$, and the $H_z$ of the electromagnetic field is an even mode decoupling from the radiation mode due to the plane wave is an odd mode under a normal incident around the $z$ axis, which is called the symmetric-protected BIC. Furthermore, we show the simulated transmission spectra of the nanodisk measurface as a function of the asymmetry parameters in the following two case. For the free-standing nanodisk measurface, Fig. \ref{fig2}(c) exhibits an ideal BIC with vanished linewidth around 665 nm. For the realistic case where the substrate is considered in Fig. \ref{fig2}(d), we find the ideal BIC appeared at 680 nm slightly redshift compared with the free-standing case. To further show that the substrate only shifts the resonance of the quasi-BIC but not its modal properties, we perform a transmission spectrum of $\alpha$ =0.17 (the D point of Fig. \ref{fig2}(d)), as displayed in Fig. \ref{fig2}(e). The sharp resonance can be well fitted by the Fano formula, which show that the ideal-BIC is transformed into quasi-BIC by breaking the in-plane symmetry of the unit cell with the exchange of energy. To identify the electromagnetic mode origin of the observed quasi-BIC resonance, we calculate the multipolar decomposition of the optical response under the Cartesian coordinates in Fig. \ref{fig2}(f), and it is clearly that the dominant contribution at resonance is provided by the MD, which is consistent with the analysis of the previous mentioned $H_z$ results. 

In order to facilitate the design of the self-hybridized exciton-polaritons in bulk WS$_2$ nanodisk measurface, we study the optical response of the exciton mode ($\varepsilon_b=1$) and the quasi-BIC mode ($f_0=0$) with different structure parameters. In Figs. 3(a) – 3(c), we plot the absorption spectrum of the exciton mode for different structure parameters. From these results, it can be found that the absorption of the exciton mode will increase with the height, which shows an opposite trend with the increase of the asymmetry parameter and period due to the decrease of the filling ratio. It is worthy to notice that the position of the exciton mode still locates around 620 nm. In Figs. 3(d) – 3(f), the line width of the MD quasi-BIC mode is sensitive to the increase of asymmetry parameter. The position of the MD quasi-BIC resonance shows a clear red shift in the transmission spectrums with the increase of period and height, which shows an opposite trend with the increase of the asymmetry parameter. These make it possible to manipulate the line width by adjusting the asymmetry parameter, and the resonance position through changing the period. The implementation of strong coupling in a bulk WS$_2$ metasurface needs the line width and the resonance position of the MD quasi-BIC are close to that of the exciton mode. The line width coincidence can be achieved by adjusting the asymmetry parameter, and the resonance position coincidence can be obtained by adjusting the period. Because the height has a great influence on the absorption and the line width of the exciton mode, it remains the same in the subsequent structure design. For simplification, the height is assumed as 35 nm.
\begin{figure}[htbp]
	\centering
	\includegraphics
	[scale=0.42]{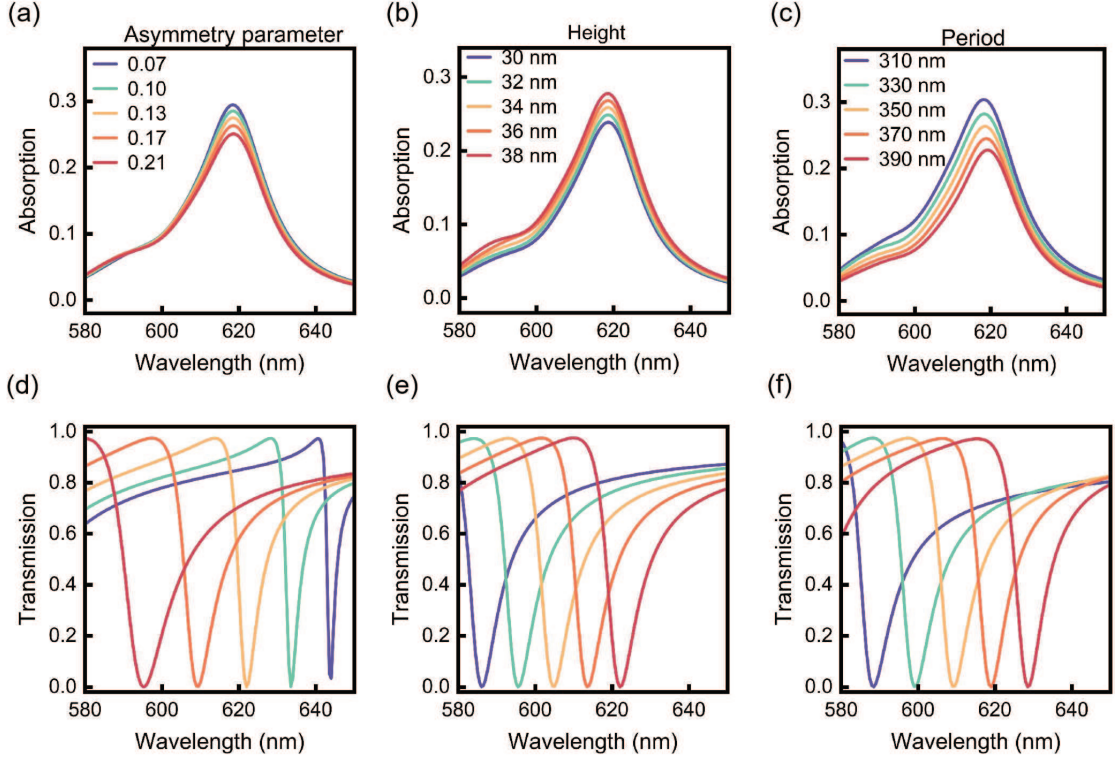}
	\caption{\label{fig3} The effect of structural parameters on the exciton mode ($\varepsilon_b=1$) and the quasi-BIC mode ($f_0=0$): (a) and (d) show the absorption spectrum of the exciton mode and the transmission spectrum of the quasi-BIC mode for different asymmetry parameters, where $P$=370 nm, $H$=35 nm. (b) and (e) show the absorption spectrums of the exciton mode and the transmission spectrums of the quasi-BIC mode for different heights, where $P$=370 nm, $\alpha$=0.17.  (c) and (f) show the absorption spectrums of the exciton mode and the transmission spectrums of the quasi-BIC mode for different periods, where $\alpha$=0.17, $H$=35 nm.}
\end{figure}

  \begin{figure}[htbp]
  	\centering
  	\includegraphics
  	[scale=0.45]{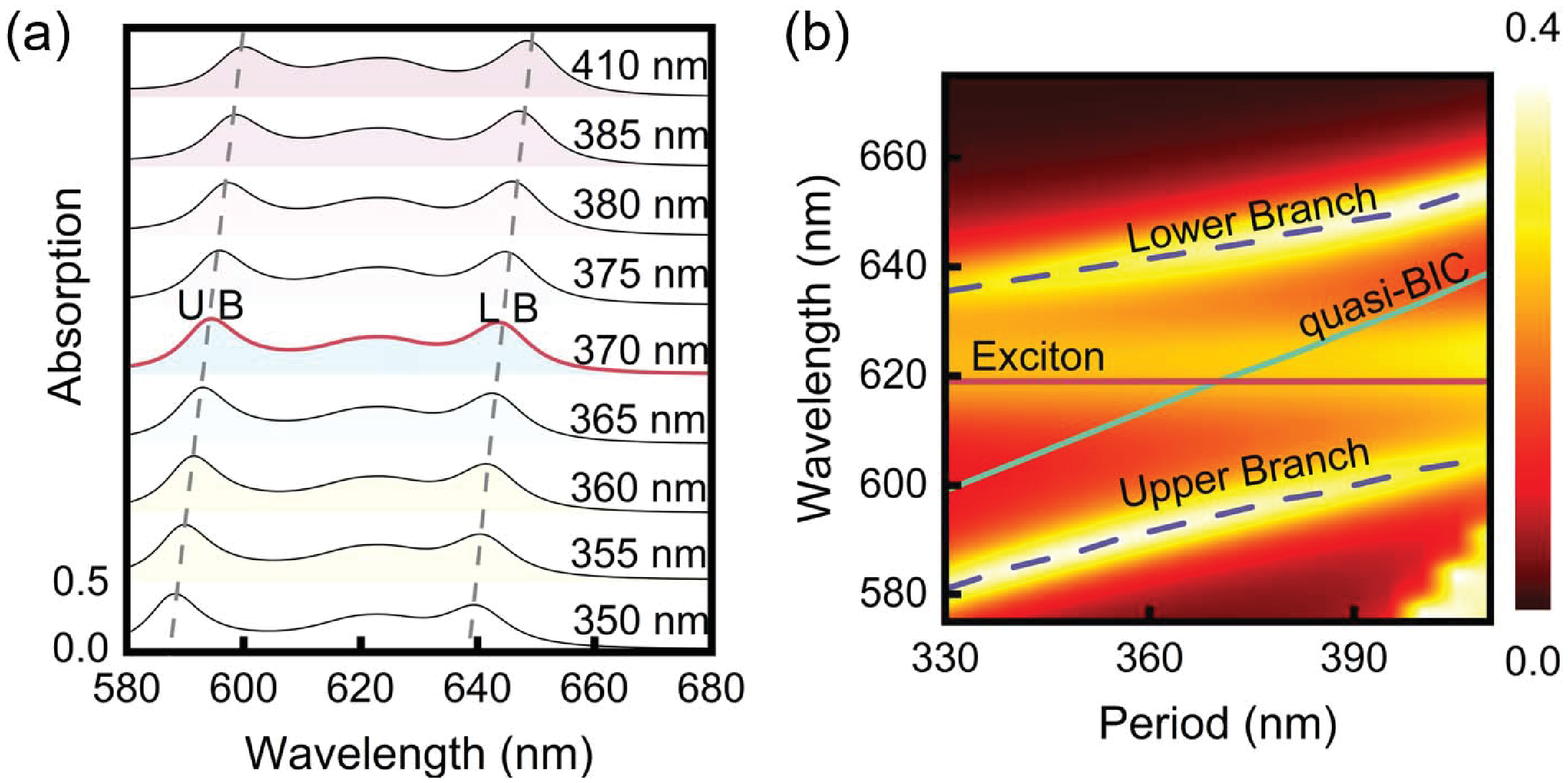}
  	\caption{\label{fig4} (a) Absorption spectrum of the bulk WS$_2$ measurface for different period $P$ at $\alpha$ =0.17, $H$=35 nm. (b) Simulated absorption contour map with the different period  $P$ at $\alpha$ =0.17, $H=35$ nm, where the red line and green line represent the individual exciton and MD quasi-BIC mode; blue dashed lines are the UB and LB, respectively.}
  \end{figure}

To realize strong coupling, we consider not only the background refractive index but also the exciton refractive index in the following  calculation. As displayed in Fig. \ref{fig4}(a), we perform a rich collection of hybridize mode absorption spectrum with the asymmetry parameter $\alpha=0.17, H=35$ nm. There are two peaks marked with UB and LB in each  absorption spectrum, showing the evidence of  strong coupling via changing the period of the nanodisk unit cell. Meanwhile,  these peaks exhibit a slight red shift owing to the increase of the effective refractive index. It is worth noting that there is another smaller peak between these two peaks in absorption spectrum,  which represents the exciton not coupled to the MD quasi-BIC mode. These results can also be observed in the anti-crossings of the 2D color map, as depicted in Fig. \ref{fig4}(b).

To describe the strong coupling scenario, we construct a Hamiltonian model based on the temporal coupled mode theory\cite{Fan_2003}. This model consists of  three modes: a MD quasi-BIC mode of the lossless cavity, a coupling involved exciton, and a surplus exciton. Thus, this system response can be written as  a complex amplitude $|a\rangle=(a_{quasi-BIC},a_{Exciton},a_{surplus})^T$, and the dynamic equation can be  given in the following step
\begin{equation}
	\frac{d|a\rangle}{dt}= \hat{H}|a\rangle+\hat{D}^T|s_+\rangle,
	\label{eq2}
\end{equation}
\begin{equation}
	|s_-\rangle= \hat{C}|s_+\rangle+\hat{D}|a\rangle,
	\label{eq3}
\end{equation}
where $|s_+\rangle=[s_{1+}$,$s_{2+}]^T$ and $|s_-\rangle=[s_{1-}$,$s_{2-}]^T$ describe amplitudes of incoming and outgoing waves. $\hat{C}$ is the direct scattering matrix.
radiate coupling matrix $\hat{D}$=
$\begin{pmatrix}
	\sqrt \gamma_{B}  &  0   &  0\\ 
	\sqrt \gamma_{B}  &  0   &  0\\
\end{pmatrix}$ and $\hat{H}$ is the Hamiltonian of the coupled system:
\begin{equation}
	\hat{H}=i\begin{pmatrix}
		\omega_{B}+i\gamma_{B}  &  g   &  0 \\ 
		g  &  \omega_{E}+i\gamma_{E}   &  0\\
		0  &  0   &  \omega_{surplus}+i\gamma_{surplus}\\
	\end{pmatrix},
	\label{eq4}
\end{equation}
where $\omega_{B}$ and $\gamma_{B}$ are the frequency and half-widths of the quasi-BIC mode, respectivly. $\omega_{E}$ ($\omega_{surplus}$) and $\gamma_{E}$ ($\gamma_{surplus}$) represent the frequencies and half-widths of the excitons involved in coupling (surplus excitons), respectively. $g$ is the coupling strength. By calculating $\hat{H}|a\rangle=\omega |a\rangle$, the three energies of the upper branch (UB), lower branch (LB), and surplus exciton with the condition 
$\hbar\omega_{B}=\hbar\omega_{surplus}=\hbar\omega_{E}$= 2 eV  can be described as:  
\begin{equation}
	\Omega_{LB}=\frac{1}{2}(i\gamma_{B}+i\gamma_{E}+2\omega_{E}-\sqrt{4g^2-\gamma_{B}^2+2\gamma_{B}\gamma_{E}-\gamma_{E}^2}),
	\label{eq5}
\end{equation}
\begin{equation}
	\Omega_{UB}=\frac{1}{2}(i\gamma_{B}+i\gamma_{E}+2\omega_{E}+\sqrt{4g^2-\gamma_{B}^2+2\gamma_{B}\gamma_{E}-\gamma_{E}^2}),
	\label{eq6}
\end{equation}
\begin{equation}
	\Omega_{surplus}=i\gamma_{surplus}+\omega_{E}.
	\label{eq7}
\end{equation}
With these expressions, the Rabi splitting can be given by
\begin{equation}
	\hbar\Omega=\hbar\Omega_{UB}-\hbar\Omega_{LB}=\sqrt{4g^2-(\gamma_{B}-\gamma_{E})^2}.
	\label{eq8}
\end{equation}

Taking the red line in Fig. \ref{fig4}(a) as an example, the UB is around 595 nm and the LB is around 644 nm, and the Rabi splitting $\hbar\Omega=\hbar\Omega_{UB}-\hbar\Omega_{LB}$$\approx$159 meV, which is larger than the conventional Rabi splitting formed in monolayer TMDCs integrated on dielectric resonators\cite{Qin2021,Qin2022}. The exciton mode involving in coupling can be defined as the difference between the total excitons $\gamma_{total}$ and the surplus excitons $\gamma_{surplus}$. By calculating the half value of full width at a half maximum, we can obtain the $\hbar\gamma_{total}=38.1 $ meV and the $\hbar\gamma_{surplus}=22.4$ meV, thus $\hbar\gamma_{E}=\hbar\gamma_{total}-\hbar\gamma_{surplus}=15.7$ meV.  Due to the $\hbar\gamma_{B}=14.3$ meV can be extracted by fitting the numerical simulation transmission spectrum $T$ with Fano model shown in Fig. \ref{fig2} (e), then we can obtain the coupling strength $\hbar g=79.5$ meV from the Eq. (8), which satisfies the condition of strong coupling $g>\frac{\gamma_B-\gamma_E}{2}$.

 \begin{figure}[htbp]
 	\centering
 	\includegraphics
 	[scale=0.6]{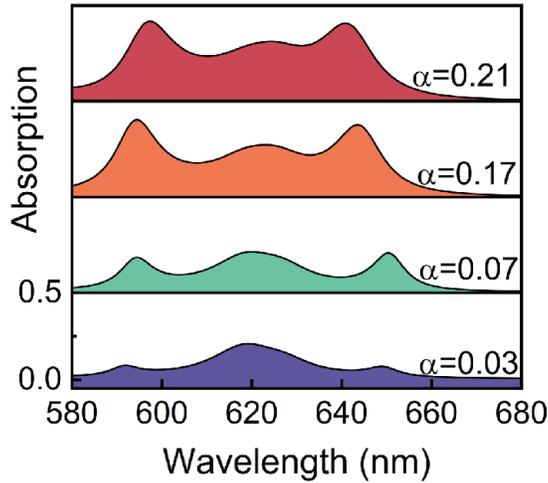}
 	\caption{\label{fig5} Absorption spectrums of the bulk WS$_2$ measurface for different asymmetry parameters under the energy detuning is zero.}
 \end{figure}

In addition, we can tune the interaction between excitons and quasi-BIC from weak coupling to strong coupling by adjusting the asymmetry parameters, in the following discussion. Considering an extreme example that the $\gamma_{B}$ as a line width is vanishingly small, which represents the coupling strength $\hbar g=0$, the absorption will exhibit a pure exciton mode owing to the ideal BIC is a dark mode that not couples with any exciton mode. This can be also achieved from Eq.(\ref{eq5}), Eq.(\ref{eq6}), and Eq.(\ref{eq7}), when $\hbar\gamma_{B}=0$ meV and $\hbar g=0$ meV, these equations can be deduced: $\Omega_{LB}=\omega_{E}$, $\Omega_{UB}=i\gamma_{E}+\omega_{E}$, $\Omega_{surplus}=i\gamma_{surplus}+\omega_{E}$. This means the UB and LB degenerate into ideal BIC and exciton modes, respectively, resulting a pure exciton mode without the splitting peaks in the absorption spectrum. when we set a smaller asymmetry parameter for BIC, it will couple to the exciton mode, enabling the system first enter the weak coupling region.  This can be captured by the blue area with $\alpha=0.03$ in Fig. \ref{fig5}. It is worth noticing that the absorption of UB and LB is close to zero, which means the number of photons is much less than excitons, leading to the surplus excitons paly a dominant role in the absorption spectrum. As the $\gamma_{B}$ becomes larger,  we can obtain three solutions from the Eq.(\ref{eq5}) and Eq.(\ref{eq6}), which means the system enters the strong coupling region with the apperacance of the specral splitting. This can be proved by the other cases with $\alpha$=0.07, 0.17, and 0.21 in Fig. \ref{fig5}. We can see that with the increase of the asymmetry parameter, the absorption of the UB and LB increases synchronously, while the absorption of the surplus exciton mode decreases significantly. This is because the number of photons that can participate in the coupling are becoming larger, but are still less than the total excitons, resulting a peak of the surplus exciton mode always exists in the absorption spectrum.

\section{\label{sec4}Conclusions}

In conclusion, we realize the strong coupling between MD quasi-BIC and the excitons occurs within the proposed bulk WS$_2$ measurface with the Rabi splitting up to 159 meV. Further we describe the coupling scene by constructing a Hamiltonian model containing the surplus excitons based on the temporal coupled mode theory, enabling a discussion of the coupling state evolution through considering surplus excitons. Our work could deepen understanding of surplus exciton mode and the absorption manipulation in the strong light-matter interaction, which provide a theoretical contributions to the designable and novel exciton-polariton devices.
  
\begin{acknowledgments}	
This work is supported by the National Natural Science Foundation of China (Grants No. 11947065, No. 12064025), the Natural Science Foundation of Jiangxi Province (Grants No. 20202BAB211007, No. 20212ACB202006), the Interdisciplinary Innovation Fund of Nanchang University (Grant No. 20199166-27060003), the Open Project of Shandong Provincial Key Laboratory of Optics and Photonic Devices (Grant No. K202102), the Major Discipline Academic and Technical Leaders Training Program of Jiangxi Province  (Grant No. 20204BCJ22012).

\end{acknowledgments}


%

\end{document}